\documentclass[aps,showpacs,superscriptaddress,preprint]{revtex4}%
\usepackage{amsfonts}
\usepackage{amsmath}
\usepackage{amssymb}
\usepackage{graphicx}%
\begin{document}
\title{Adsorption and diffusion of H$_{2}$O molecule on the Be(0001) surface: A
density-functional theory study}
\author{Shuang-Xi Wang}
\affiliation{State Key Laboratory for Superlattices and Microstructures, Institute of
Semiconductors, Chinese Academy of Sciences, P. O. Box 912, Beijing 100083,
People's Republic of China}
\affiliation{Tsinghua, People's Republic of China}
\affiliation{LCP, Institute of Applied Physics and Computational Mathematics, P.O. Box
8009, Beijing 100088, People's Republic of China}
\author{Peng Zhang}
\affiliation{Department of Nuclear Science and Technology, Xi'an Jiaotong University, Xi'an
710049, People's Republic of China}
\author{Jian Zhao}
\affiliation{State Key Laboratory for Geomechanics and deep underground engineering, China
University of Mining and Technology, Beijing 100083, People's Republic of China}
\author{Shu-Shen Li}
\affiliation{State Key Laboratory for Superlattices and Microstructures, Institute of
Semiconductors, Chinese Academy of Sciences, P. O. Box 912, Beijing 100083,
People's Republic of China}
\author{Ping Zhang}
\thanks{Corresponding author. zhang\_ping@iapcm.ac.cn}
\affiliation{LCP, Institute of Applied Physics and Computational Mathematics, P.O. Box
8009, Beijing 100088, People's Republic of China}


\begin{abstract}
Using first-principles calculations, we systematically study the adsorption
behavior of a single molecular H$_{2}$O on the Be(0001) surface. We find that
the favored molecular adsorption site is the top site with an adsorption
energy of about 0.3 eV, together with the detailed electronic structure
analysis, suggesting a weak binding strength of the H$_{2}$O/Be(0001) surface.
The adsorption interaction is mainly contributed by the overlapping between
the $s$ and $p_{z}$ states of the top-layer Be atom and the molecular orbitals
1$b_{1}$ and 3$a_{1}$ of H$_{2}$O. The activation energy for H$_{2}$O
diffusion on the surface is about 0.3 eV. Meanwhile, our study indicates that
no dissociation state exists for the H$_{2}$O/Be(0001) surface.

\end{abstract}
\maketitle

Water adsorption on metal surfaces has gained a lot of interest in a variety
of phenomena such as heterogeneous catalysis and corrosion of materials
\cite{Thiel1987,Henderson2002}. As a result these systems have been
intensively investigated by various experimental and theoretical techniques.
With respect to the substrates, most of the recent studies have focused on the
transition metal surfaces, such as Cu(100) \cite{Brosseau1993,Sanwu2004},
Fe(100) \cite{Hung1991,Jung2010}, and Pd(100) \cite{Lloyd1986,Jibiao2007}.
Although ambiguities exist, by early density-functional theory (DFT)
calculations, theoretically a flat-lying configuration on the top site of
transition metal surfaces has been established by some sophisticated studies
\cite{Michaelides2003,Sheng2004,Carrasco2009}, arguing that the molecular
orbitals (MOs) of water, mainly 1$b_{1}$, dominate the water-surface
interaction, by coupling with atomic $d$ orbital of the transition metal
surfaces. In comparison with the vast studies on the adsorption properties of
H$_{2}$O on transition metal surfaces, however, for simple metals which are
lack of $d$ states, very few researches \cite{Michaelides2004} have been
reported up to now. This long-term overlook should be altered by considering
the fact that there is now an increasing practical demand for a thorough
understanding of the water structures on some specific simples metals.

Motivated by this observation, in the present paper we use first-principles
calculations to investigate the adsorption properties of H$_{2}$O on the
Be(0001) surface. The reason why we choose beryllium as the prototype for
simple metals is as follows: (i) beryllium has vast technological applications
due to its high melting point and low weight. One of its important usage is as
a plasma facing material in experimental nuclear fusion reactors
\cite{Diez1990}. As the first wall, the beryllium can adsorb the residual
gases in the plasma vessel, improving the vacuum cleanliness
\cite{Argentina2000}. Water is the main residual gas in the ultrahigh vacuum
(UHV) vessels of the fusion reactors, so it is highly meaningful to study the
adsorption of water molecule at beryllium surfaces; (ii) besides and
prominently, different from the bulk, beryllium surfaces have a large
directional $s$ and $p$ electronic distributions around the Fermi energy and
thus may display unique interaction with the adsorbed H$_{2}$O molecules.

Through analysis of projected density of states (PDOS) and charge density
difference, we obtain the adsorption properties of H$_{2}$O on the Be(0001)
surface. It is found that the H$_{2}$O molecule prefers to adsorb on the
surface top site in a fairly flat-lying configuration with a weak binding
strength. The adsorption interaction is mainly characterized by the
overlapping between the $s$ and $p_{z}$ states of the top-layer Be atom and
the MOs 1$b_{1}$ and 3$a_{1}$ of the water molecule. We notice that different
from transition metal surfaces, the MO 3$a_{1}$ plays an important role in the
interaction between H$_{2}$O and Be(0001). The diffusion energetics of H$_{2}%
$O across the Be(0001) surface is also studied, which gives an activation
barrier of $\sim$0.3 eV.

Our calculations are performed within DFT using the Vienna \textit{ab-initio}
simulation package (VASP) \cite{Kresse1996}. The PW91 \cite{Perdew1992}
generalized gradient approximation and the projector-augmented wave potential
\cite{Kresse1999} are employed to describe the exchange-correlation energy and
the electron-ion interaction, respectively. The cutoff energy for the plane
wave expansion is set to 400 eV. The Be(0001) surface is modeled by a slab
composing of five atomic layers and a vacuum region of 20 \AA . A 3 $\times$ 3
supercell, in which each monolayer contains nine Be atoms, is adopted in the
study of the H$_{2}$O adsorption. The water is placed on one side of the slab
only and a dipole correction \cite{Bengtsson1999} is applied to compensate for
the induced dipole moment. During our calculations, the bottom two atomic
layers of the Be surface are fixed, and other Be atoms as well as the H$_{2}$O
molecule are free to relax until the forces on the ions are less than 0.02
eV/\AA . Integration over the Brillouin zone is done using the Monkhorst-Pack
scheme \cite{Monkhorst1976} with 7 $\times$ 7 $\times$ 1 grid points. And a
Fermi broadening \cite{Weinert1992} of 0.1 eV is chosen to smear the
occupation of the bands around the Fermi energy ($E_{F}$) by a finite-$T$
Fermi function and extrapolating to $T$ = 0 K.

The calculation of the energy barriers for the water diffusion processes is
performed using the nudged elastic band (NEB) method \cite{Jonsson1998}, which
is a method for calculating the minimum energy path between two known minimum
energy sites, by introducing a number of \textquotedblleft
images\textquotedblright\ along the diffusion path. Then the energy barrier is
determined by relaxing the atomic positions of each image in the direction
perpendicular to the path connecting the images, until the force converges. In
present work, the diffusion path is modeled using seven images, two of which
include the minimum energy sites as initial and final positions, with five
linearly interpolated, intermediate images between them.

The structural and energetic parameters of the free water molecule
are calculated within a box with the same size of the adsorbed
systems. The optimized geometry for free H$_{2}$O gives a bond
length of 0.97 \AA ~ and a bond angle of 104.4$^{\circ}$, consistent
with the experimental values of 0.96 \AA ~ and 104.4$^{\circ}$
\cite{Eisenberg1969}. The calculated lattice constant of bulk Be
($a$, $c$) are 2.26 \AA ~and 3.58 \AA , respectively, in good
agreement with the experimental measurements of 2.29 \AA ~and 3.588
\AA \ \cite{Wachowicz2001}.

\begin{figure}[ptb]
\begin{center}
\includegraphics[width=0.5\linewidth]{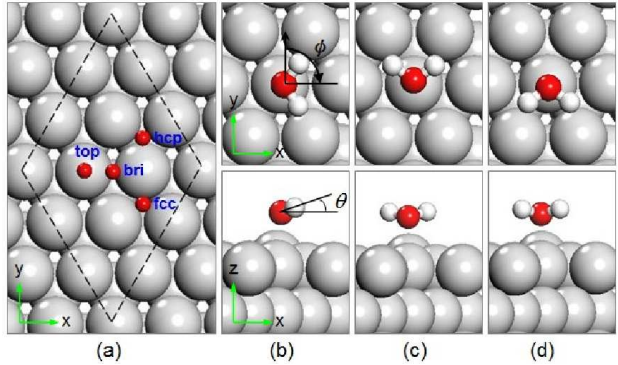}
\end{center}
\caption{(Color online) (a) The structure of the p($3\times3$) surface cell of
Be (0001), and three on-surface adsorption sites. The red balls denote the
initial positions of O atoms in the adsorption picture. (b)-(d) Top view
(upper panels) and side view (lower panels) of the optimized structures of
three most stable adsorption states of H$_{2}$O/Be(0001) surface, i.e.,
top-$x$, top-$y1$ and top-$y2$, respectively. Gray, red and white balls denote
Be, O, and H atoms, respectively.}%
\label{fig1}%
\end{figure}

As depicted in Fig. \ref{fig1}(a), there are four high-symmetry sites on the
Be(0001) surface, respectively the top, bridge (bri), hcp and fcc hollow
sites. H$_{2}$O adsorption at all four high symmetry adsorption sites on
Be(0001) surface is considered. The O atom of water is initially placed on the
precise high-symmetry sites with various orientations of water with respect to
the substrate. We find that there exist locally stable adsorption states on
the top site of Be(0001), where the H$_{2}$O molecules lie fairly flat on the
surface, labeled by employing the notations top-$x$, $y1$ and $y2$,
respectively. The structural and energetic details of the molecular states are
illustrated and summarized in Figs. \ref{fig1}(b)-(d) and Table \ref{table1},
respectively. In Fig. \ref{fig1}(b) we define two angles $\phi$ and $\theta$.
$\phi$ represents the azimuthal angle of H$_{2}$O with respect to the surface,
and $\theta$ represents the tilt angle between the H$_{2}$O molecular dipole
plane and the surface. The adsorption energy of the system is calculated as
follows:
\begin{equation}
E_{\mathrm{ad}}=E_{\mathrm{H_{2}O/Be(0001)}}-E_{\mathrm{H_{2}O}}%
-E_{\mathrm{Be(0001)}}, \label{Ead}%
\end{equation}
where $E_{\mathrm{H_{2}O}}$, $E_{\mathrm{Be(0001)}}$, and $E_{\mathrm{H_{2}%
O/Be(0001)}}$ are the total energies of the H$_{2}$O molecule, the clean Be
surface, and the adsorption system respectively. According to this definition,
a negative value of $E_{\mathrm{ad}}$ indicates that the adsorption is
exothermic (stable) with respect to a free H$_{2}$O molecule and a positive
value indicates endothermic (unstable) reaction.

\begin{table}[ptbh]
\caption{Calculated structural parameters, adsorption energy for a water
molecule on Be(0001) surface. E$_{a}$ (eV) represents the adsorption energy.
$\Phi$ (eV) represents the work function. $z_{\text{O}}$ (\AA ) represents the
vertical height of the O atom from the surface. $d_{\text{O-H}}$ (\AA )
represents the bond length between the O and H atoms. $\theta$ ($^{\circ}$)
represents the tilt angle between the H$_{2}$O molecular dipole plane and the
surface. $\alpha_{\text{H-O-H}}$ ($^{\circ}$) represents the H-O-H bond
angle.}%
\label{table1}%
\begin{tabular}
[c]{ccccccc}\hline\hline
\ Site & E$_{a}$ & $\Phi$ & $z_{\text{O}}$ & $d_{\text{O-H}}$ & $\theta$ &
$\alpha_{\text{H-O-H}}$\\\hline
\ top-$x$ & -0.300 & 3.75 & 1.79 & 0.98 & 21.6 & 106.3\\
\ top-$y1$ & -0.303 & 3.75 & 1.79 & 0.98 & 20.7 & 106.4\\
\ top-$y2$ & -0.300 & 3.75 & 1.79 & 0.98 & 20.6 & 106.4\\\hline\hline
\  &  &  &  &  &  &
\end{tabular}
\end{table}

From Table \ref{table1}, we can clearly see that at these stable adsorption
sites, the work function 3.75 eV is much smaller than the clean Be(0001)
surface (5.41 eV), implying an observable charge redistribution between the
adsorbate water and the surface Be atoms. Take the adsorption site top-$y1$
for example, the O-H bond length 0.98 \AA ~is almost identical to 0.97 \AA ~of
free H$_{2}$O molecule, but the H-O-H bond angle 106.4$^{\circ}$ is larger
than that of free H$_{2}$O. The tilt angle is 20.7$^{\circ}$, which is in line
with the value of about 20$^{\circ}$ of the H$_{2}$O/Al(100) surface
\cite{Michaelides2004}, but differing from that ($\sim$10$^{\circ}$) of the
H$_{2}$O/transition metal surfaces, where the top site is also the most stable
state. The adsorption energy $-$0.303 eV indicates a weak molecule-surface
interaction, and the molecular state may diffuse easily with a low barrier,
passing through the bridge sites as a transition state, which will be
discussed in the latter part of this paper. More interestingly, we find that
for a H$_{2}$O molecule to reach the adsorption state, there does not exist
any energy barrier, which means that H$_{2}$O can be adsorbed on the Be(0001)
surface spontaneously. This is quite different from the adsorption of other
molecules on the Be(0001) surface such as O$_{2}$ \cite{Zhang2009} and CO
\cite{Wang2009}, which need to overcome energy barriers.

\begin{figure}[ptb]
\begin{center}
\includegraphics[width=0.5\linewidth]{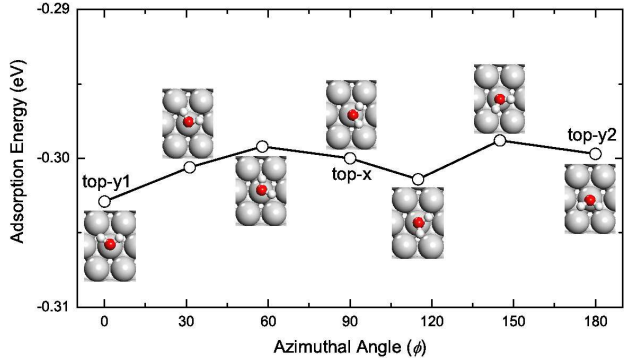}
\end{center}
\caption{(Color online) Calculated adsorption energy of H$_{2}$O as a function
of the azimuthal angle at the top site. The insets show the structures adopted
in the calculations.}%
\label{fig2}%
\end{figure}

As presented in Table \ref{table1}, the energetic differences among these
three adsorption states on top site are very tiny. For further illustration,
we investigate the azimuthal orientation of adsorbed H$_{2}$O, which is shown
in Fig. \ref{fig2}. It can be seen that the favored orientation for H$_{2}$O
at the top site is top-$y1$. We can attribute this to the symmetry of the
adsorption structure. The Be atom at the second layer (on the hcp site) of the
substrate might also be partially responsible for this tiny azimuthal
orientation preference. Nevertheless, it is apparent that the azimuthal
rotation is facile, therefore the adsorption of H$_{2}$O on the top site is
insensitive to the azimuthal orientation. In addition, a considerably less
stable adsorption state, where the water molecule is adsorbed on the top site
in an upright configuration with the O atom down, is found by our
calculations. the adsorption energy of this state is about 0.19 eV, which is
also insensitive to the azimuthal orientation. In the following discussion, we
will focus our attention on the the most stable flat-lying adsorption sites on
the top site.

\begin{figure}[ptb]
\begin{center}
\includegraphics[width=0.5\linewidth]{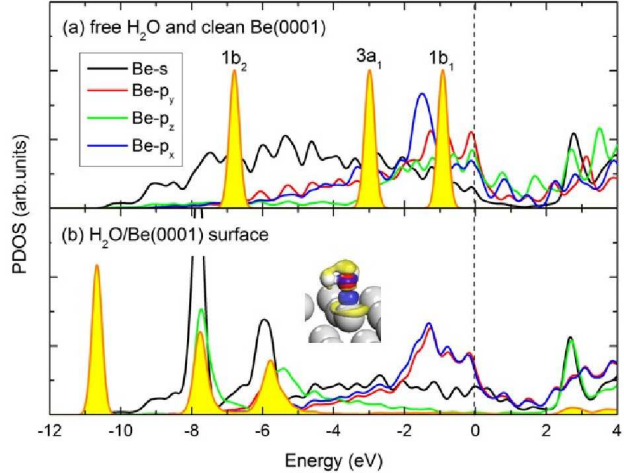}
\end{center}
\caption{(Color online) The PDOS for the H$_{2}$O molecule and the top-layer
Be atom bonded to H$_{2}$O, at the stable adsorption site top-$y1$: (a) free
H$_{2}$O and the clean Be(0001) surface, (b) the molecular state of H$_{2}%
$O/Be(0001) surface, where the inset shows the 3D electron density difference,
with the isosurface value set at $\pm$0.025 \textit{e}/\AA $^{3}$. The area
filled with yellow color represents molecular orbital of H$_{2}$O. The Fermi
energy is set to zero.}%
\label{fig3}%
\end{figure}

In order to gain more insights into the precise nature of the
chemisorbed molecular state, the electronic PDOS of the H$_{2}$O
molecule and the topmost Be layer are calculated. As a typical
example, here we plot in Fig. \ref{fig3} the PDOS for the most
stable adsorption configuration of top-$y1$. For comparison, the
PDOS of the free H$_{2}$O molecule and clean Be(0001) surface are
also shown in Fig. \ref{fig3}. Three-dimensional (3D) electron
density difference $\Delta\rho(\mathbf{r})$, which is obtained by
subtracting the
electron densities of noninteracting component systems, $\rho^{\text{Be(0001)}%
}(\mathbf{r})+\rho^{\text{H$_{2}$O}}(\mathbf{r})$, from the density
$\rho(\mathbf{r})$ of the H$_{2}$O/Be(0001) surface, while retaining the
atomic positions of the component systems at the same location as in H$_{2}%
$O/Be(0001), is also shown in the inset of Fig. \ref{fig3}(b). MOs 2$a_{1}$
and 1$b_{2}$ of water (not shown here) are far below the Fermi energy and thus
remain intact in water-metal interaction. Here we consider only three MOs of
water below the Fermi energy, i.e., 1$b_{2}$, 3$a_{1}$ and 1$b_{1}$. One can
clearly see from Fig. \ref{fig3}(a) that after adsorption, these three MOs
shift down in energy 3.8 eV, 4.8 eV and 4.9 eV, respectively. Compared to free
water molecule, the distortion of the adsorbed water, i.e., the effect of
water geometry, is pretty small (see Table \ref{table1}). Hence, these shifts
of energy can be attributed to the interactions with the substrate. In
addition to energy, it is noticeable that the MOs of adsorbed water are
broaden apparently for 1$b_{1}$ and 3$a_{1}$, which are known to have an
oxygen lone-pair character perpendicular to the molecular plane, and a mixture
of partial lone-pair character parallel to the molecular plane and partial O-H
bonding character, respectively \cite{Thiel1987}. Furthermore, despite little
change in the PDOS near the surface Fermi energy, the water adsorption
introduces new peaks for both $s$ and $p_{z}$ states of the surface Be atom,
aligning in energy with 1$b_{1}$ and 3$a_{1}$. Especially for 3$a_{1}$, more
electronic states of the Be atom appear nearby, compared with 1$b_{1}$. This
is quite in accordance with our above-mentioned result that the tilt angle of
H$_{2}$O on the Be(0001) surface is larger than that on the transition metal
surfaces, where only the 1$b_{1}$ dominates the water-surface interaction by
coupling with atomic $d$ orbital of the surfaces. Moreover, it may also give a
reason why there exists a less stable adsorption state in an upright
configuration on the top site.

All these features indicate a covalent coupling between the adsorbed water and
the substrate. This result is substantiated by the 3D electron density
difference. We can see that there exists a large charge accumulation between
the adsorbate and substrate. As depicted by the PDOS, however, no apparent
charge transfer between water and surface Be atom can be observed. In
addition, we find that at the second layer of the surface, there exists a
small charge depletion right at the fcc site, which possibly confirms our
earlier speculation that the state top-$y1$ is most stable at the top site.

\begin{figure}[ptb]
\begin{center}
\includegraphics[width=0.5\linewidth]{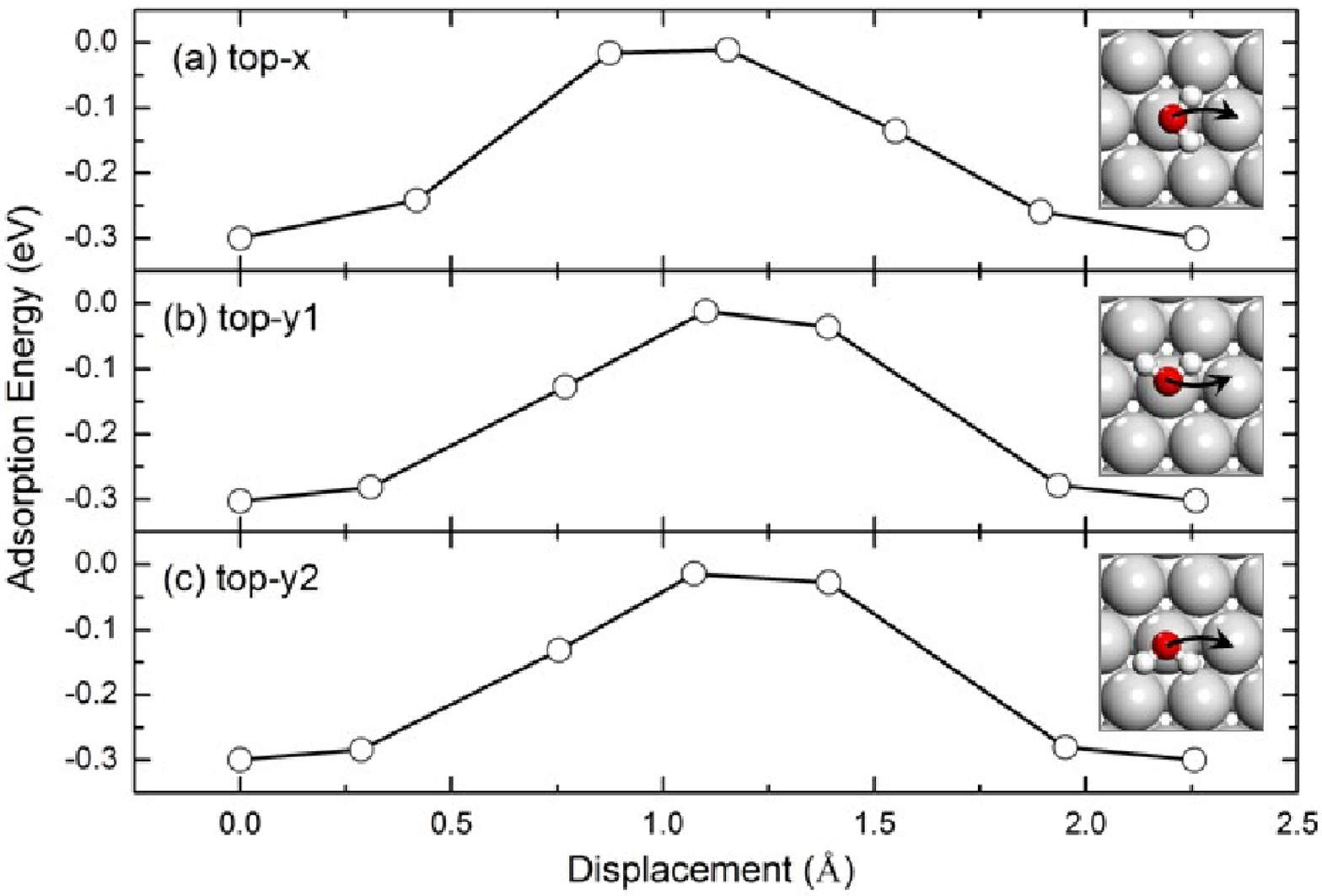}
\end{center}
\caption{Diffusion of H$_{2}$O on the top site of Be(0001) surface as a
function of the lateral displacement of O atom from its original site top-$x$
(upper panel), top-$y1$ (middle panel), and top-$y2$ (lower panel),
respectively.}%
\label{fig4}%
\end{figure}

Given a H$_{2}$O molecule at a stable adsorption site, it is interesting to
see how it can diffuse across the Be(0001) surface. Here, therefore, we
calculate the diffusion paths and energetic barriers of water on Be(0001)
surface between neighboring adsorption sites along the top-$x$, top-$y1$ and
top-$y2$ channels respectively, which are schematically shown in the insets of
Fig. \ref{fig4}. The adsorption energies as a function of the lateral
displacement of O atom are shown in Fig. \ref{fig4}. We can see that the
activation energies for H$_{2}$O diffusion along the three paths are 0.289,
0.291 and 0.286 eV, respectively. The transition state of diffusion is located
on the bridge site, which is not a stable adsorption site. The similarity of
the energy barriers between these paths implies that it does not have a strong
preference for a particular direction in which to diffuse, consistent with our
previous analysis illustrated in Fig. \ref{fig2}. By assuming the attempt
frequency 10$^{13}$ of the adsorbate, the energy barrier corresponds to
diffusion temperature of about 113 K, suggesting that the diffusion can occur
under room temperature on the H$_{2}$O/Be(0001) surface.

Finally let us discuss the possibility of dissociation of H$_{2}$O on the
Be(0001) surface. Experimentally, it was found that beryllium can be oxidated
by the water vapor \cite{Zalkind2005}. Basically, since the binding strength
of H$_{2}$O/Be(0001) system is weak, hence one anticipates that it is not
prone to dissociate for H$_{2}$O on the ideal Be(0001) surface under low
temperature. This is confirmed by our first-principles static as well as
molecular dynamics simulations. We find that the hollow hcp and fcc sites are
stable both for H and OH species on the Be(0001) surface. When we put the H
and OH species simultaneously at two nearest-neighbor hollow sites, however,
it turns out that they spontaneously combine together into H$_{2}$O molecule
finally without any barrier encountered. This is the same for the
next-nearest-neighbor condition. The MD simulations are performed using the
Verlet algorithm with a time step of 1 fs within the micro canonical ensemble.
The O atom of the H$_{2}$O molecule is initially set to be 4 \AA ~away from
the metal surface. The substrate atoms are initially at rest, while the
initial kinetic energy of H$_{2}$O is set to be 0.6 eV, with the initial
velocity towards the substrate. In the same conditions it has been found that
O$_{2}$ molecule would dissociate finally \cite{Yang2010}. Here for H$_{2}$O,
nevertheless, it rapidly reaches a stable state that is similar to the
adsorption state found above. Consequently, it proves that no dissociation
state exists for the H$_{2}$O/Be(0001) surface under room temperature. We
speculate that the experimentally observed surface oxidation of beryllium by
water vapor is closely related to the surface roughness or defects.

In conclusion, we have systematically studied the adsorption behavior of
H$_{2}$O on the Be(0001) surface by using first-principles DFT method. It has
been found that the water molecule prefers to adsorb on the top site by lying
fairly flat on the surface, insensitive to the azimuthal orientation and with
a weak binding strength. It has been shown that the MO 3$a_{1}$, as well as
1$b_{1}$, plays an important role in H$_{2}$O-Be interaction by overlapping
with the $s$ and $p_{z}$ states of the underlying beryllium atom. The
diffusion energetics of H$_{2}$O on the top site across the Be(0001) surface
has been calculated to display an energy barrier of about 0.3 eV. Moreover, no
dissociation states of water molecule are found for the ideal Be(0001) surface.

\end{document}